\documentclass{bmcart}

\usepackage[utf8]{inputenc} 

\usepackage{graphicx,subfigure}
\usepackage{xcolor}
\usepackage{soul}
\usepackage{tabularx, booktabs} 
\usepackage{color}
\usepackage{amsmath}
\usepackage[inline]{enumitem}
\usepackage{hyperref}

\newcommand{\ra}[1]{\renewcommand{\arraystretch}{#1}}

\startlocaldefs
\endlocaldefs

\begin{document}

\begin{frontmatter}

\begin{fmbox}
\dochead{Research}


\title{Using Personal Environmental Comfort Systems to Mitigate the Impact of Occupancy Prediction Errors on HVAC Performance}


\author[
   addressref={aff1},             
   corref={aff1},                       
   email={milanj@iiitd.ac.in}   
]{\inits{MJ}\fnm{Milan} \snm{Jain}}
\author[
   addressref={aff2},
   email={rachel@ee.iitm.ac.in}
]{\inits{RKK}\fnm{Rachel K} \snm{Kalaimani}}
\author[
   addressref={aff3},
   email={keshav@uwaterloo.ca}
]{\inits{SK}\fnm{Srinivasan} \snm{Keshav}}
\author[
   addressref={aff3},
   email={cath@uwaterloo.ca}
]{\inits{CR}\fnm{Catherine} \snm{Rosenberg}}


\address[id=aff1]{
  \orgname{IIIT-Delhi},                       
  \cny{India}                                 
}
\address[id=aff2]{%
  \orgname{IIT Madras},
  \cny{India}
}
\address[id=aff3]{%
  \orgname{University of Waterloo},
  \cny{Canada}
}



\end{fmbox}


\begin{abstractbox}

\begin{abstract} 
Heating, Ventilation and Air Conditioning (HVAC) consumes a significant fraction of energy in commercial buildings. Hence, the use of optimization techniques to reduce HVAC energy consumption has been widely studied. Model predictive control (MPC) is one  state of the art optimization technique for HVAC control which converts the control problem to a sequence of optimization problems, each over a finite time horizon. In a typical MPC, future system state is estimated from a model using predictions of model inputs, such as building occupancy and outside air temperature. Consequently, as prediction accuracy deteriorates, MPC performance--in terms of occupant comfort and building energy use--degrades. In this work, we use a custom-built building thermal simulator to systematically investigate the impact of occupancy prediction errors on occupant comfort and energy consumption. Our analysis shows that in our test building, as occupancy prediction error increases from 5\% to 20\% the performance of an MPC-based HVAC controller becomes worse than that of even a simple static schedule. However, when combined with a personal environmental control (PEC) system, HVAC controllers are considerably more robust to prediction errors.
Thus, we quantify the effectiveness of PECs in mitigating the impact of
forecast errors on MPC control for HVAC systems.

\end{abstract}




\end{abstractbox}
%

\end{frontmatter}



\section{Introduction}
	\label{sec:introduction}
Commercial buildings account for about one-third of global energy consumption, with HVAC (Heating Ventilating and Air-Conditioning) units being the major contributor. An HVAC typically comprises of few Air Handling Units (AHUs), which heat or cool the air to a specified setpoint temperature, and Variable Air Volume (VAV) units that control the volume of air flowing into each thermal zone. In most commercial buildings, HVAC maintains a desired \emph{set point} temperature during working hours (9 AM to 6 PM) and a \emph{set back} temperature during non-working hours. Unfortunately, given the stochastic nature of building occupancy, a static schedule either leads to energy wastage or occupant discomfort~\cite{dawson2013boss,erickson2013poem}.

HVAC energy optimization is an active area of research. In the literature, studies have proposed several control strategies which broadly fall into two categories: \emph{reactive}~\cite{jain2016non, gyalistras2010use} and \emph{predictive}~\cite{jain2017portable+, oldewurtel2010energy}. In a reactive controller, AHUs and VAVs respond to measured occupancy in a zone. Here, occupancy is measured using motion, $CO_2$ sensors, or by monitoring building's WiFi infrastructure~\cite{trivedi2017ischedule}. Since buildings typically take some time to respond to control inputs, better performance can be obtained using predictive control strategy where the controller selects the optimal trajectory of set points for a finite time horizon~\cite{jain2018energy}. Of the predictive control techniques, perhaps the best-known approach is Model Predictive Control (MPC)~\cite{garcia1989model}.

In a typical MPC, a known building thermal model estimates the future system state using forecasts of model inputs, such as building occupancy and outside air temperature. However, the effectiveness of this approach depends on the accuracy of the predictions. As prediction accuracy deteriorates, MPC performance - in terms of occupant comfort and building energy use - degrades and may get even worse than conventional techniques. In recent work, Oldewurtel et al.~\cite{oldewurtel2011stochastic} extensively studied the influence of errors in weather forecast on HVAC energy consumption and occupants' comfort and quantified the impact of mis-predictions. However, the work neither addressed errors in occupancy prediction nor studied the ways to mitigate the influence of prediction errors.

In this paper, we address this gap. We study the influence of occupancy errors on MPC performance using a custom-built building simulator. We also model and analyze the impact of personal environmental control system (PEC) in the presence of prediction errors. A PEC could be an off-the-shelf desktop fan or a heater to provide individual thermal comfort~\cite{brager2015evolving}. We find that PEC when used with model predictive control, can reduce both - the variability in energy consumption and the occupants' discomfort. 

Our contributions are as follows:
\begin{enumerate}
\item We present the design and development of a building thermal simulator that models conventional schedule-based, reactive occupancy-based, and predictive MPC-based HVAC controllers.
\item We extend the MPC-based control strategy proposed by Kalaimani et al.~\cite{Rac17} and allow PEC to react between any two consecutive states of the system.
\item We quantify the impact of occupancy prediction errors on two MPC-based control strategies - with and without PEC. For analysis, we use occupancy data from forty-five volunteers over three months and simulations of a test building in both heating and cooling seasons.
\item It is important that occupancy forecast errors are realistic; thus, we propose a method to systematically introduce realistic occupancy errors into MPC predictions using real-world occupancy data. 
\end{enumerate}

The rest of the paper is organized as follows. Section~\ref{sec:related} discusses the literature and studies conducted in the past. In Section~\ref{sec:background}, we outline the control strategies studied in the paper. In Section~\ref{sec:architecture}, we present the detailed architecture and design of the thermal simulator followed by detailed analysis in Section~\ref{sec:evaluation}. In Section~\ref{sec:discussions}, we discuss several limitations and possible future directions of the study and conclude the paper. 

\section{Related Work}
	\label{sec:related}
\subsection{Central HVAC Controllers}

In the past, researchers have extensively studied the optimization of HVAC controllers to minimize the aggregate energy consumption and maximize user comfort~\cite{jain2016data, jain2014pacman}. Agarwal et al.~\cite{agarwal2011duty} studied aggressive duty cycling of HVAC based on occupancy patterns within the building. Lu et al.~\cite{lu2010smart} proposed a smart thermostat to automate HVAC control by sensing occupancy and sleeping patterns in residential buildings. The occupancy-based control allows buildings to operate outside of comfort regimes when unoccupied, thus reducing energy usage~\cite{erickson2011observe}. Henceforth, several other studies also explored the use of occupancy information to optimize the HVAC energy operations~\cite{erickson2013poem,balaji2013sentinel,erickson2011observe,Nest,aswani2012reducing,iyengar2015iprogram,kleiminger2014smart,koehler2013therml,scott2011preheat,yang2012living}. However, centralized HVAC controllers divide a building into thermal zones comprising of private and shared spaces. Within each zone, these control strategies maintain ASHRAE standard while assuming each zone as either occupied or unoccupied; thus, ignoring individual comfort requirements.

\subsection{Personal Environmental Control}

For personalized comfort, studies proposed to use personal environmental control systems (PECs), especially in  shared spaces~\cite{brager2015evolving,bauman1998field,zhang2010comfort,gao2013spot,gao2013optimal,rabbani2016spot}.  Unlike conventional centrally-controlled HVAC system, where people share the same set point temperature~\cite{dear2013progress, jain2017decision}, PEC systems can meet the comfort requirements of all occupants, albeit at the cost of additional energy expenditure. Kalaimani et al.~\cite{Rac17} merged PEC with model predictive control to further minimize the HVAC energy consumption and maximize the user comfort. 

Though advanced predictive control strategies (such as MPC) have the potential to optimize HVAC operations significantly, none of the studies mentioned above quantify the influence of the prediction errors on the energy consumption of HVAC and on the occupants' comfort.

\subsection{Error Analysis}
Oldewurtel et al.~\cite{oldewurtel2011stochastic} studied the influence of errors in weather forecast on MPC-controlled HVAC operations, and their results indicate that the quality of weather predictions highly correlates with the performance of the model predictive controller. However, the study only focused on prediction errors in the weather forecast and the evaluation was limited to ``pure" MPC-based HVAC controller. Given that occupancy prediction is also an input to MPC, it is essential to analyze the influence of occupancy prediction errors on HVAC operations. Besides, the study~\cite{oldewurtel2011stochastic} is limited to HVAC and does not incorporate the impact of PECs in satisfying the comfort requirements of occupants.

In this paper, we extend the work in~\cite{oldewurtel2011stochastic} and in \cite{Rac17} by first analyzing the effect of prediction errors in occupancy and later exploring the benefits of PECs in mitigating (or minimizing) the influence of prediction errors on HVAC operations. Our study indicates that predictive control strategies make HVAC operations highly unreliable. High variability has discouraged building managers to use advanced HVAC control strategies, and thus, they have continued using conventional HVAC controllers. 

\section{HVAC Control Strategies}
	\label{sec:background}
In a typical commercial building, spaces are either private (such as offices) or shared (such as cafeteria, corridor), and a set of private and shared spaces constitutes a zone. Within each zone, there exists a VAV unit that takes air from AHU at a particular temperature~($u(t)$) and supplies it across the rooms at a specific rate~($v^{ij}{(t)}$) to maintain the room temperature close to the set point temperature. Here, $j$ indicates the room number in the $i^{th}$ zone of the building. To ensure a consistent supply of fresh air, AHU recirculates only a limited amount of used air~($r{(t)}$) and ejects the remaining air in the open environment.

Defined in Table~\ref{table:hvac_cvars}, $u(t)$, $v^{ij}{(t)}$, and $r{(t)}$ are key HVAC control parameters and their values are typically decided by the control strategy. In this section, we discuss the four control strategies, implemented to analyze the influence of occupancy prediction errors on HVAC operations. The first two methods are non-predictive, and building managers widely use these strategies in commercial buildings today; when employed, the HVAC operations are independent of prediction errors. The last two are MPC-based control strategies. In the paper, we use non-predictive control strategies as the baseline strategies for predictive control strategies when occupancy prediction is not perfect.

\subsection{Schedule-based control}
In a schedule-based control of HVAC, the building manager starts the HVAC at a fixed time in the morning and shuts it down in the evening (typically $9~AM$ to $6~PM$). On any day, AHU supplies air at a static temperature which is chosen based on the season (summer/winter), and the set point temperature does not vary within a day. Based on ASHRAE standards\footnote{American Society of Heating, Refrigeration and Air-Conditioning - a global organisation that publishes standards and guidelines related to HVAC.}, we set the supply air temperature ($u^{(t)}$) to $15^{\circ}$C for summers and $20^{\circ}$C for winters. For both seasons, the ratio of reuse air ($r$) and rate of flow of supply air ($v^{(t)}$) is constant at $0.8$ and $0.236~m^3/s$, respectively. The approach is naive but widely used by building managers in commercial buildings.

\subsection{Reactive control}
In reactive control strategies, VAV cools or heats the space only if people are present in the corresponding VAV zone. In the past, studies have suggested several direct and indirect HVAC control strategies to estimate occupancy; we use the occupancy data and implement the strategy proposed by Ardakanian et al.~\cite{ardakanian2016non} for benchmarking. 

\begin{table}[h!]
\centering
\caption{List of HVAC control variables}
\ra{1.5}
    \begin{tabular}{@{}lm{6.5cm}rl@{}}\toprule[0.3ex]
        \textbf{Symbol}       	& \textbf{Description}                                                      & \textbf{Unit}             \\
        \hline
        $u{(t)}$  				&   Supply air temperature at time $t$                                      & $^\circ$C                 \\
        $v^{ij}{(t)}$  			&   Rate of flow of supply air in room $j$ of a VAV zone $i$ at time $t$    & $m^3/s$                   \\
        \bottomrule[0.3ex]
    \end{tabular}%
    \label{table:hvac_cvars}
\end{table}

\subsection{Model Predictive Control}
Model predictive control (MPC) is a recent approach for HVAC where controller can compute the room temperature over a finite time horizon~\cite{mpcsurvey}. Typically, a thermal model using occupancy estimates and weather forecast determines the future system state over a time horizon. In our implementation of MPC, we used Equation~\ref{eq:tm_hvac} as the thermal model that considers the influence of HVAC, atmospheric temperature, heating load by the occupants, and other heating or cooling loads present in the room~\cite{riederer2002room}. Table~\ref{table:param_desc_mpc} lists all symbols of the thermal model and their default values.
\begin{equation}
	\begin{split}
	\frac{T^{ij}{(t+1)} - T^{ij}{(t)}}{\tau} \times C^{ij} &= \frac{\rho\sigma}{n_r^i} \times v^{ij}{(t)} \times (u{(t)} - T^{ij}{(t)}) \\ & + \alpha_{ex}^{ij} \times (T_{ex}{(t)} - T^{ij}{(t)}) \\ &+ (Q_{oc}^{ij} + Q_{ap}^{ij}) \times O^{ij}{(t)}
	\end{split}
	\label{eq:tm_hvac}
\end{equation}
For the time horizon, the controller computes $u{(t)}$, $v^{ij}{(t)}$, and $r{(t)}$, by solving an optimization problem using the current state of the system, with an objective to minimize the total energy consumption (Equation~\ref{eq:min_power_hvac}).
\begin{equation}
	\begin{split}
		Po{(t)} &= V{(t)} \times \eta_{h} \times (u{(t)} - T_{cu}{(t)}) + V{(t)} \times V{(t)} \times \eta_{\mathit{f}}\\ &+ V{(t)} \times \eta_{c} \times (T_{mx}{(t)} - T_{cu}{(t)})
	\end{split}
	\label{eq:min_power_hvac}
\end{equation}

where,
\begin{equation}
	\begin{split}
	V{(t)} = \sum_{i=1}^{n_z}\sum_{j=1}^{n_r^i}{v^{ij}{(t)}}
	\end{split}
	\label{eq:vt}
\end{equation}
$V{(t)}$ depicts the total air supplied across all the rooms within a building, $\eta_h$ and $\eta_c$ indicate the efficiency of the heating and cooling unit, respectively. $T_{cu}(t)$ and $T_{mx}(t)$ denote the temperature of air coming from the cooling and mixing unit, respectively. $\eta_f$ is the efficiency of the VAV fan which is supplying air to the room. Details about the power consumed by the supply fan can be found in Rabbani et al.~\cite{rabbani2016spot}.

The optimization problem constraints the comfort index to remain within the specified bounds to ensure user comfort. In this study, we use widely used metric PMV - Predicted Mean Vote, to measure user comfort~\cite{fanger1973assessment}. Other constraints include time-scale limitations, thermal dynamics (Equations \ref{eq:tm_hvac}), and constraints dictated by the system setup (such as thermal comfort and HVAC operation should remain within a desired range). In this paper, we implemented MPC with two time-scales where the controller updates the supply air temperature every hour and the supply air volume every 10 minutes. The above time-scales are typically determined by the physical limitation of an HVAC unit. For more details about this
specific formulation of the optimization problem, 
please refer to Kalaimani et al.~\cite{kalkesros16}.

\begin{table}[ht!]
\caption{List of symbols used in the thermal model}
\ra{1.5}
    \begin{tabular}{@{}lm{6cm}rl@{}}\toprule[0.3ex]
        \textbf{Symbol}     & \textbf{Description}       & \textbf{Default}       & \textbf{Unit}       \\
        \hline
        $\rho$              &   Density of air                                                          & $1.204$   & $kg/m^3$      \\
        $\sigma$            &   Specific heat of air                                                    & $1.003$   & $kJ/(kg.K)$   \\
        $\tau$              &   Sampling interval                                                       & $-$       & $s$           \\
        $n_z$               &   Total number of VAV zones in a building                                 & $-$       & $-$           \\
        $n_r^i$             &   Total number of rooms in VAV zone $i$                                   & $-$       & $-$           \\
        $O^{ij}{(t)}$       &   Occupancy in room $j$ of zone $i$ at time $t$                           & $-$       & $-$           \\
        $T^{ij}{(t)}$       &   Temperature in room $j$ of zone $i$ at time $t$ due to HVAC             & $-$       & $^\circ C$    \\
        $T_{ex}{(t)}$       &   External temperature at time $t$
                            & $-$       & $^\circ C$    \\
        $C^{ij}$            &	 Thermal capacity of room $j$ in zone $i$ 
                            & $2000$    & $kJ/K$        \\
        $\alpha_{ex}^{ij}$  &   Heat transfer coefficient between outside and room $j$ in zone $i$     & $0.048$   & $kJ/(K.s)$    \\
        $Q_{ap}^{ij}$       &   Heat load due to heating/cooling equipments in room $j$ of zone $i$     & $0.1$     & $kW$          \\
        $Q_{oc}^{ij}$       &   Heat load due to occupant in  room $j$ of zone $i$ 
                            & $0.1$     & $kW$          \\
        \bottomrule[0.3ex]
    \end{tabular}%
    \label{table:param_desc_mpc}
\end{table}

\subsection{MPC with Personal Environment Controller}
Recently, Kalaimani et al.~\cite{Rac17} proposed a hybrid HVAC controller and combined MPC with a personal environmental control system. In the study, \cite{Rac17} used SPOT - an off-the-shelf desktop fan/heater with local temperature sensing and a computer-controlled actuator to provide individual thermal comfort. Assuming perfect prediction of occupancy and outside temperature, the study shows that combining MPC with SPOT is effective in reducing the total energy consumption by choosing appropriate thermal setbacks during the intervals of sparse occupancy.

At the time of partial occupancy, HVAC runs at a base temperature which is slightly higher (in summers) or lower (in winters) than the desired temperature. Equation~\ref{eq:tm_hvac} depicts the base temperature (due to HVAC) which depends on HVAC, external weather conditions, and occupants within the space. 
\begin{equation}
	\begin{split}
	\frac{T_{hv}^{ij}{(t+1)} - T_{hv}^{ij}{(t)}}{\tau} \times C^{ij} &= \frac{\rho\sigma}{n_r^i} \times v^{ij}{(t)} \times (u^{ij}{(t)} - T_{hv}^{ij}{(t)}) \\ & + \alpha_{ex}^{ij} \times (T_{ex}{(t)} - T_{hv}^{ij}{(t)})  + Q_{ap}^{ij} \times O^{ij}{(t)}
	\end{split}
	\label{eq:tm_thvac}
\end{equation}

In the proposed approach, we assume that room is divided into two regions: 
\textit{occupied }- the part of the room where the occupant is present; and 
\textit{unoccupied} - the other part of the room.

In the occupied region, SPOT provides the offset comfort to attain the comfort requirements of the occupant. In Equation~{\ref{eq:tm_thvac}}, we show how the room temperature changes when taking into account  the impact
of HVAC, heat exchange with the outside, external weather conditions, and other heating/cooling loads present in the room. In Equation~\ref{eq:tm_delta}, we then calculate the change in temperature due to SPOT, occupant, and other heat exchanging load present in the room, followed by temperature in the occupied part of the room in Equation~\ref{eq:tm_toc}. On the other hand, in the unoccupied portion (Equation~\ref{eq:tm_tun}), both SPOT and the occupants \textit{indirectly} influence the room temperature due to thermal coupling between the two zones, modeled by the heat transfer coefficient $\alpha_{in}$.Table~\ref{table:param_desc_spot} lists the new notations used in the extended model.

\begin{equation}
	\begin{split}
	\frac{\Delta_{oc}^{ij}{(t+1)} - \Delta_{oc}^{ij}{(t)}}{\tau} \times C_{oc}^{ij} &= Q_{oc}^{ij} \times O^{ij}{(t)} + Q_{he}^{ij} \times S_{he}^{ij}{(t)}\\& - \alpha_{in}^{ij} \times \Delta_{oc}^{ij}{(t)}
	\end{split}
	\label{eq:tm_delta}
\end{equation}
\begin{equation}
	T_{oc}^{ij}{(t+1)} = T_{hv}^{ij}{(t+1)} + \Delta_{oc}^{ij}{(t+1)}
	\label{eq:tm_toc}
\end{equation}
\begin{equation}
	T_{un}^{ij}{(t+1)} = T_{hv}^{ij}{(t+1)} + \frac{\tau \times \alpha_{in}}{C^{ij} - C_{oc}^{ij}} \times \Delta_{oc}^{ij}{(t)}
	\label{eq:tm_tun}
\end{equation}

The revised objective function has additional parameters $S_{f}$ for fan and $S_{he}$ for heater (Equation~\ref{eq:min_power_spot}). A fan consumes negligible power, thus 
the objective function only considers the power consumption of SPOT's heater. 

\begin{equation}
	\begin{split}
	    Po{(t)} &= V{(t)} \times \eta_{h} \times (u{(t)} - T_{cu}{(t)}) + V{(t)} \times V{(t)} \times \eta_{\mathit{f}}\\ &+ V{(t)} \times \eta_{c} \times (T_{mx}{(t)} - T_{cu}{(t)}) + \sum_{i=1}^{n_z}\sum_{j=1}^{n_r^i}{S_{he}^{ij}{(t)}}
	\end{split}
	\label{eq:min_power_spot}
\end{equation}

The controller determines HVAC control parameters (Table~\ref{table:hvac_cvars}) on a 10-minute timescale and in between, fan/heater (of SPOT) reacts to occupancy every 30 seconds. By doing so, SPOT assists the controller in regulating the discomfort that might arise due to mis-predictions; thus ensuring both - personalized comfort and minimal influence of prediction errors on HVAC operations. Next, we discuss the simulator. 

\begin{table}[h!]
\caption{Notations used in the revised thermal model}
\ra{1.5}
    \begin{tabular}{@{}lm{5cm}rl@{}}\toprule[0.3ex]
        \textbf{Symbol}         & \textbf{Description}                                                                          & \textbf{Default}  & \textbf{Unit}       \\
        \hline
        $T_{hv}^{ij}{(t)}$    	&   Temperature in room $j$ of zone $i$ at time $t$ due to HVAC                                 & $-$               & $^\circ C$ \\
        $\Delta_{oc}^{ij}{(t)}$	&   Change in temperature of occupied region of room $j$ in zone $i$ at time $t$                 & $-$               & $^\circ C$ \\
        $T_{oc}^{ij}{(t)}$		&   Temperature in occupied region of room $j$ in zone $i$ at time $t$                        & $-$               & $^\circ C$ \\
        $T_{un}^{ij}{(t)}$		&   Temperature in unoccupied region of room $j$ in zone $i$ at time $t$                        & $-$               & $^\circ C$ \\
        $C_{oc}^{ij}$         	&   Thermal capacity of occupied region of room $j$ in zone $i$                                 & $200$             & $kJ/K$    \\
        $\alpha_{in}$       	&   Heat transfer coefficient between occupied and unoccupied regions of room $j$ in zone $i$  & $0.1425$          & $kJ/(K.s)$ \\
        $Q_{he}^{ij}$        	&   Heat load due to SPOT in  room $j$ of zone $i$ 
                                                    & $0.7$             & $kW$
        \\
        \bottomrule[0.3ex]
    \end{tabular}%
    \label{table:param_desc_spot}
\end{table}

\section{Simulator Software Architecture}
	\label{sec:architecture}
\begin{figure}[h!]
	\includegraphics[width=0.8\linewidth]{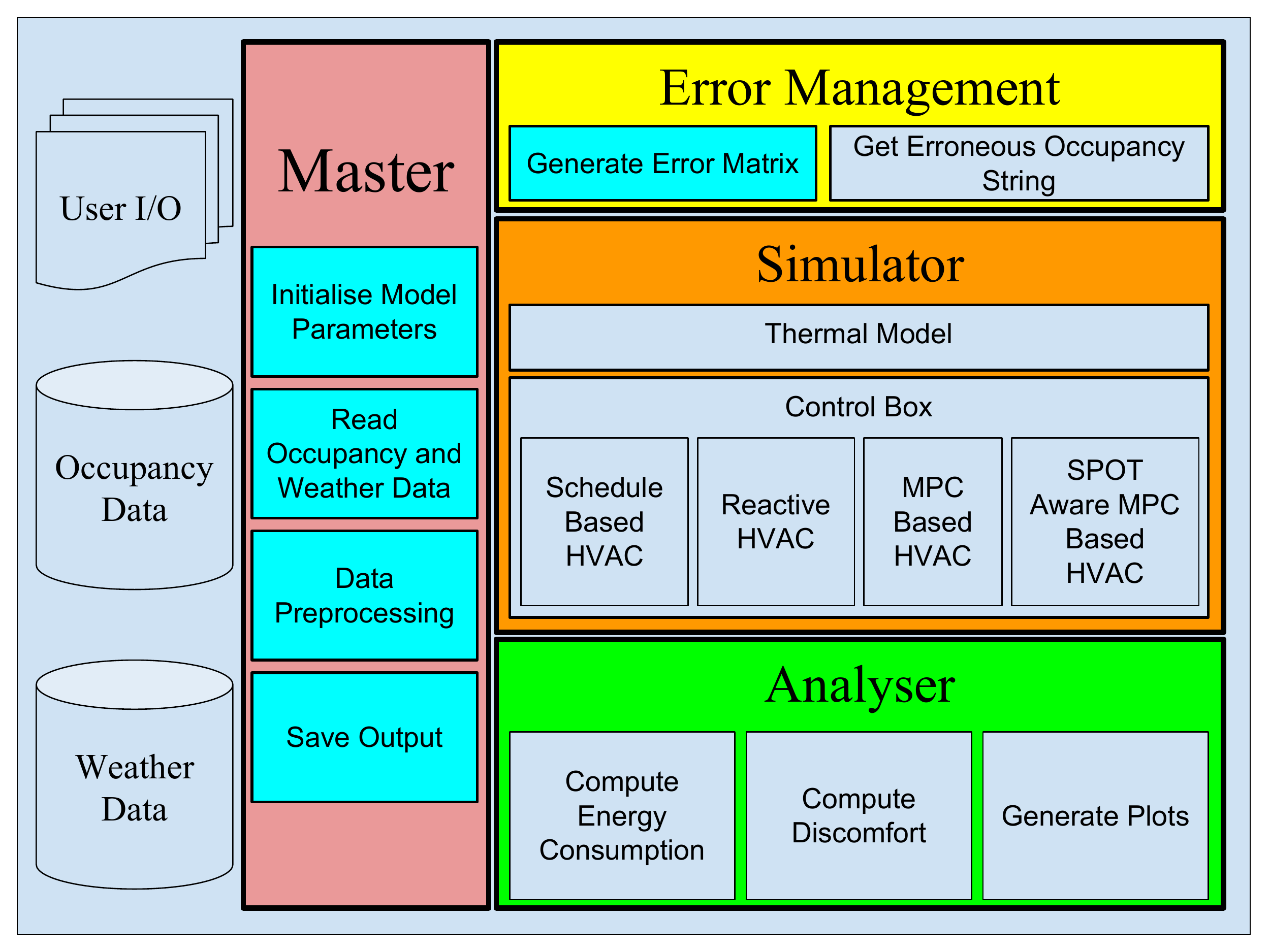}
    \caption{\csentence{Architecture.}
      ThermalSim is a lightweight C/C++ based building simulation platform that focuses on analysing the influence of prediction errors on HVAC operations.}
    \label{fig:architecture}
\end{figure}

To evaluate the impact of forecast errors on the different HVAC controllers, we built a custom open source
thermal simulator called ThermalSim~\cite{jain2017thermalsim}. ThermalSim is a lightweight C/C++ based simulation platform, whose focus is to study the influence of prediction errors on HVAC operations. 
Figure~\ref{fig:architecture} outlines the architecture of ThermalSim. It consists of four major modules: 
\begin{enumerate}
	\item Master - to handle data I/O and preprocessing,
	\item Error Management -  to inject \textit{unbiased} errors in the occupancy streams,
	\item Simulator - to simulate room temperature for a given thermal model and control logic, and
	\item Analyser - to compute energy consumption, occupant comfort, and analyze simulated data streams.
\end{enumerate}
In the current version, Simulator module incorporates AMPL~\cite{AMPL} -- an algebraic modeling language for the mathematical programming -- to compute the control parameters. 

\subsection{Master Module}
\begin{figure}[h!]
    \includegraphics[width=\linewidth]{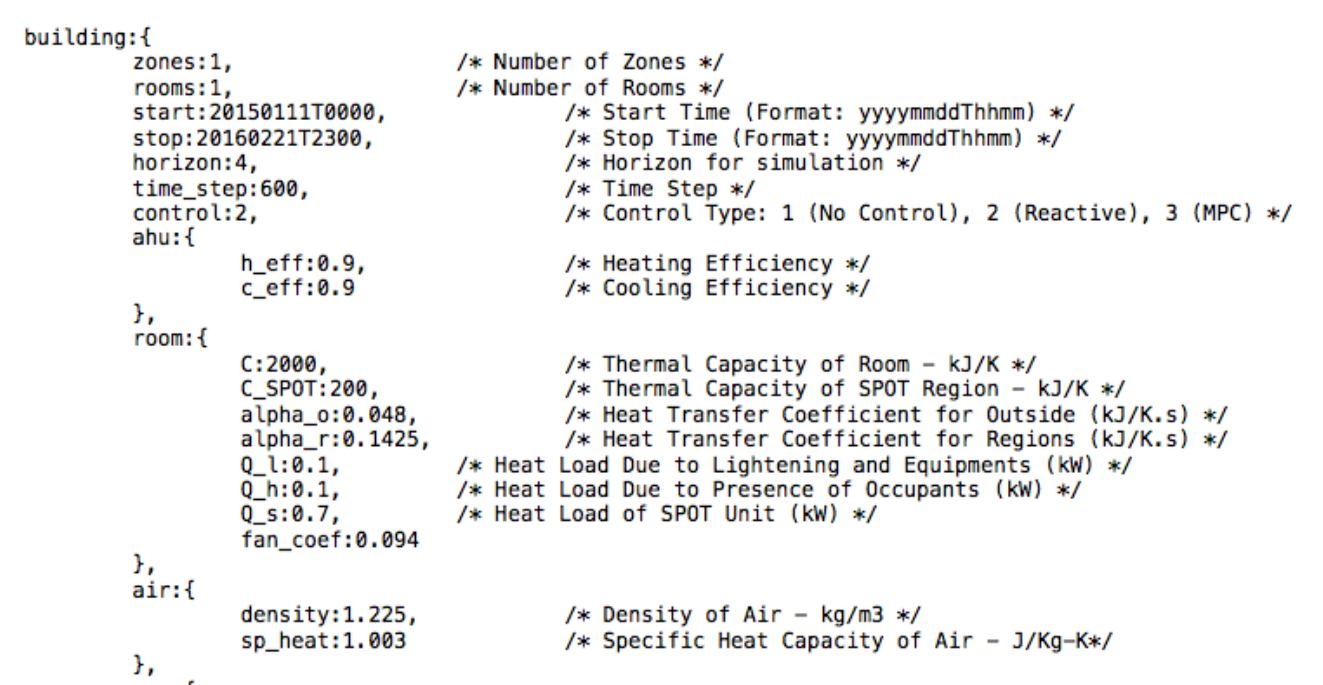}
    \caption{\csentence{Input format for ThermalSim.}}
    \label{fig:input}
\end{figure}

The Master module takes as input historical weather and occupancy data in CSV (Comma Separated Values) format, a user-generated description of the building, and simulation control parameters (Figure~\ref{fig:input}) including  start and stop time of the simulation, parameters of the thermal model, control strategy, among others. Before executing the simulations, the Master module pre-processes the data, and after completion saves the output of simulation in the CSV format.

\subsection{Modeling Occupancy Prediction Errors}
ThermalSim represents occupancy data for a day as a string of consecutive 0's (for unoccupied workspaces) and 1's (for occupied spaces). We consider only two states of occupancy because a majority of occupancy prediction algorithms use occupancy as a two-state variable. We call this string an \emph{occupancy string}. The length of a single occupancy string depends upon the sampling rate of the occupancy data. Data sampled every ten minutes will generate an occupancy string of length 144 characters, and if the sampling rate is thirty seconds, the string will be 2880 characters long.

\subsubsection{Error Matrix}
It is important that occupancy forecast errors be realistic.
For example, it does not make sense to randomly flip occupancy states,
since this may result in forecasting occupancy during the middle of the night, which is very unlikely.
Our key insight is that a likely outcome of an errored forecast is to forecast \textit{another valid occupancy string}, with the observation that the higher the error rate, the larger the distance, in an appropriate
metric space, between the true and the errored strings.

We use the following approach: For a dataset with $n$ occupancy strings, each cell of an \emph{error matrix} depicts the Hamming Distance between any two occupancy strings -- the number of mismatching characters~\cite{hamming1950error}. To normalize, we divide value in each cell by the length of occupancy string. The \emph{error matrix} is a 
symmetric matrix of size $n^2$ which helps in systematically injecting unbiased errors in the occupancy data. 

To illustrate, consider a scenario where we want to analyze different control strategies with 10\% prediction error in the occupancy data. The error management module will refer \emph{error matrix} for an occupancy string which is closest to the day of analysis. We term the selected occupancy string as the \emph{reference} string. The module will then look into the \emph{error matrix} to find all those strings that have 10\% error as compared to the \emph{reference string} and randomly select one. We call the selected one an {erroneous} string. If the day (\emph{reference} string) was 30\% occupied, then the occupancy in the \emph{erroneous} string may fall anywhere in between 20\%-40\%. 

\begin{figure}[h!]
	\includegraphics[width=0.8\linewidth]{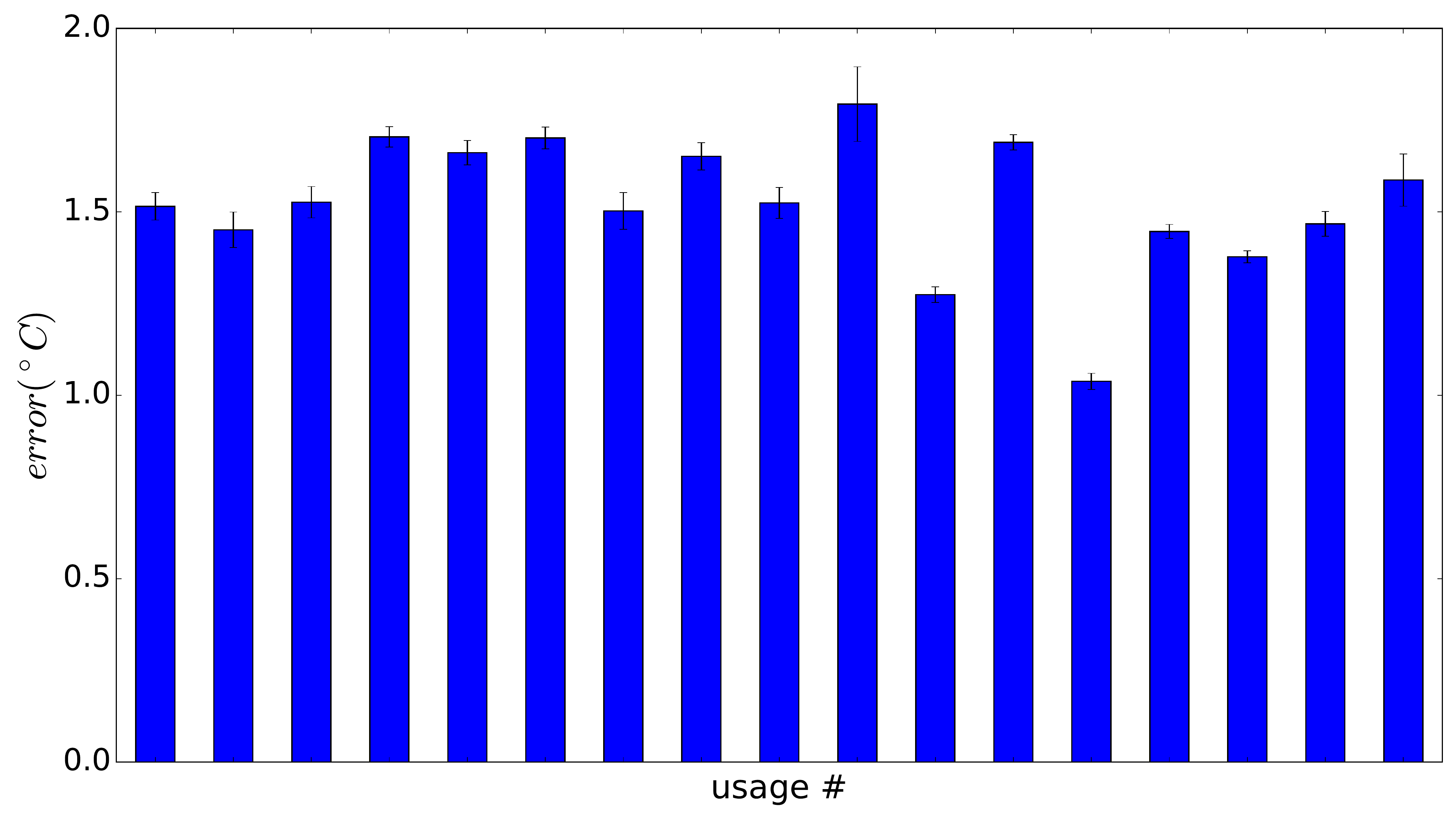}
    \caption{\csentence{Simulation Error.}
      ThermalSim can simulate daily room temperature with an RMSE of $1.52^\circ C (\sigma = 0.18^\circ C)$.}
    \label{fig:simulation_error}
\end{figure}

\begin{figure}[h!]
	\includegraphics[width=0.6\linewidth]{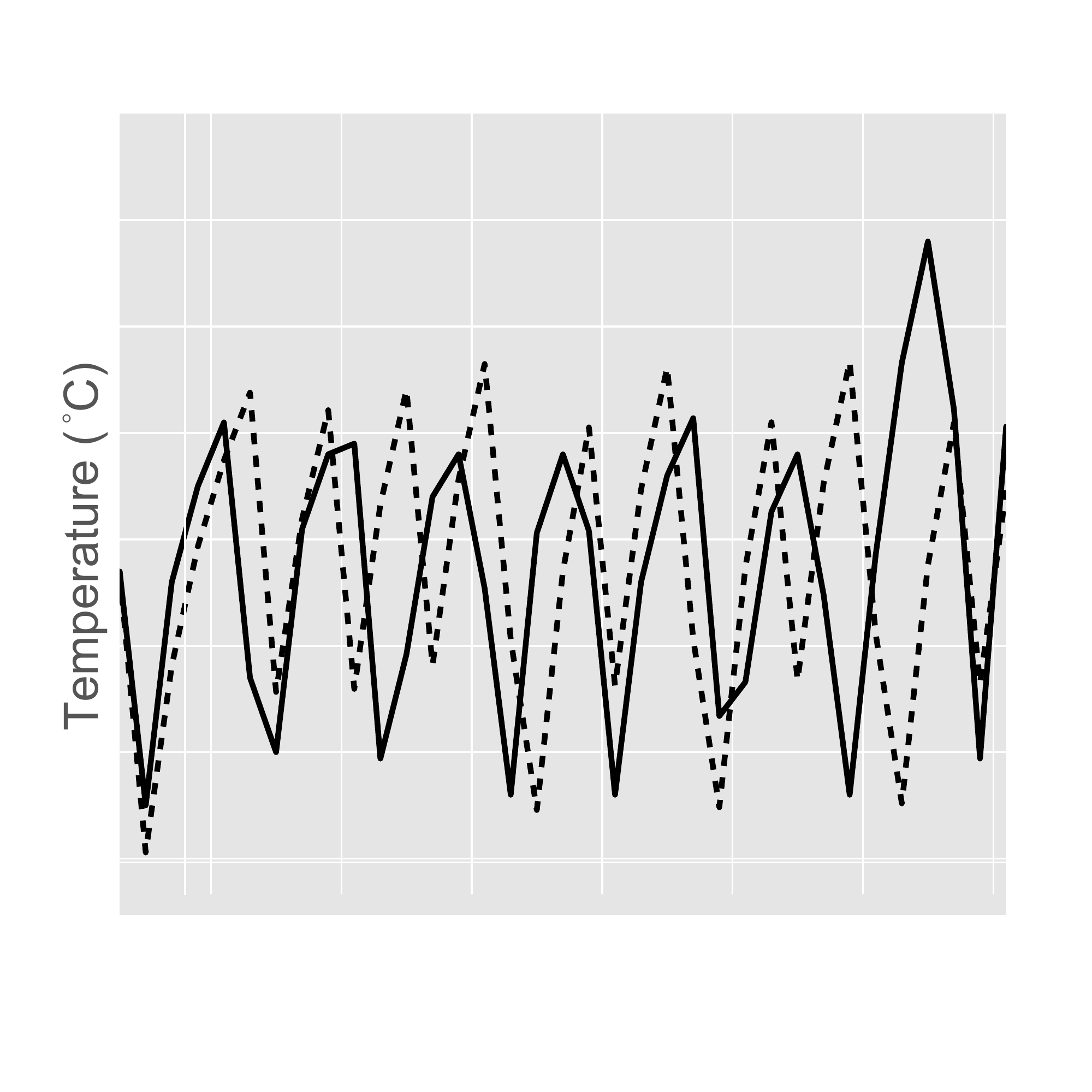}
    \caption{\csentence{Simulation Results.}
      The hard line indicates the actual room temperature and dotted line indicates the predicted room temperature. }
    \label{fig:simulation_results}
\end{figure}

\subsection{Simulator}
The simulator module takes input from the master and error management modules to simulate the room temperature. It comprises two major blocks - 
\begin{enumerate*}
	\item thermal model - depicts various thermal interactions occurring within a room, and
	\item control module - to compute the control parameters. 
\end{enumerate*} 
In the current version, we have implemented two thermal models - 
\begin{enumerate*}
	\item single region - no partition exists within a room (Equation~\ref{eq:tm_hvac}), and
	\item two regions - the occupied area is separated from the unoccupied portion by a thin layer of air (Equations~\ref{eq:tm_thvac}-\ref{eq:tm_tun}). 
\end{enumerate*} 
As discussed in Section~\ref{sec:background}, we have implemented four HVAC controllers in ThermalSim - 
\begin{enumerate*}
	\item schedule-based,
	\item reactive,
	\item model predictive control (no SPOT device present), and 
	\item SPOT-aware model predictive control.
\end{enumerate*} 
In the rest of the paper, we will use NS as an acronym for No-SPOT model predictive control and SA for SPOT-Aware MPC. 

\subsubsection*{Simulator Validation}
To quantify the accuracy of \emph{ThermalSim} in simulating room temperature, from a room in residential apartment, we collected temperature data for $17$ days and carried out leave-p-out cross validation with $p=5$. In such an approach, we validate the model on $p$ observations and use the remaining observations for training. We used a non-linear solver whose objective was to minimize the residual between predicted and actual room temperature. The simulator tunes following model parameters - 
\begin{enumerate}
    \item thermal capacity of the room ($C$),
    \item heat transfer coefficient between outside and room ($\alpha_{ex}$),
    \item coefficient of heating/cooling ($\rho \sigma$)    
    \item heat load due to occupants ($Q_{ac}$), and
    \item heat load due to heating/cooling appliances ($Q_{ac}$). 
\end{enumerate}
Our analysis (in Figure~\ref{fig:simulation_error}) indicates that \emph{ThermalSim} can simulate the daily room temperature with an RMSE (Root Mean Square Error) of $1.52^\circ C (\sigma = 0.18^\circ C)$. Figure~\ref{fig:simulation_results} depicts the average (solid line) and predicted (dashed line) room temperature. Note that though the predicted room temperature follows the pattern of actual room temperature, it fails to align perfectly. Though misalignment does increase the RMSE at some time instances, 
we found that it has little overall impact on total energy consumption and occupants' comfort. 

\subsection{Metrics}

\subsubsection{Energy Consumption} Equation~\ref{eq:energy} computes the total energy consumption of a building for a day. Here, $Po(t)$ denotes the power consumption of HVAC and other heating/cooling devices, $\tau$ is the sampling rate, and $n_t$ is the number of daily samples. 
	
	\begin{equation}
		E = \sum_{t=0}^{n_t} Po{(t)} \times \frac{\tau}{3600}
		\label{eq:energy}
	\end{equation}

\subsubsection{Occupant Discomfort} ThermalSim leverages Predicted Mean Vote (PMV)~\cite{ASHRAE} to estimate the comfort level of the occupants~(Equation~\ref{eq:pmv}). At a given time instant $t$, if PMV ($P^{ij}(t)$) lies within the comfort requirements ($[P_{ll}, P_{ul}]$) of an individual then we mark the room as comfortable, else uncomfortable. $D_{\%}^{ij}$ denotes the percentage of time instances in a day when the user was uncomfortable in the room. 

	\begin{equation}
		P^{ij}{(t)} = P1 \times T_{oc}^{ij}{(t)} - P2 \times v_{a}^{ij}{(t)}+ P3 \times v_{a}^{ij}{(t)} \times v_{a}^{ij}{(t)} - P4
	\label{eq:pmv}
 	\end{equation}
 	
	\begin{equation}
		D^{ij}{(t)} = max(0, P_{ll} - P^{ij}(t), P^{ij}(t) - P_{ul})
		\label{eq:dc}
	 \end{equation}
	
	\begin{equation}
		D_{\%}^{ij} = \frac{\sum_{t=0}^{n_t}[{D^{ij}{(t)} \ne 0}]}{\sum_{t=0}^{n_t}[{O^{ij}{(t)} = 1}]}
		\label{eq:dcp}
	\end{equation}

\begin{figure}[h!]
	\includegraphics[width=0.8\linewidth]{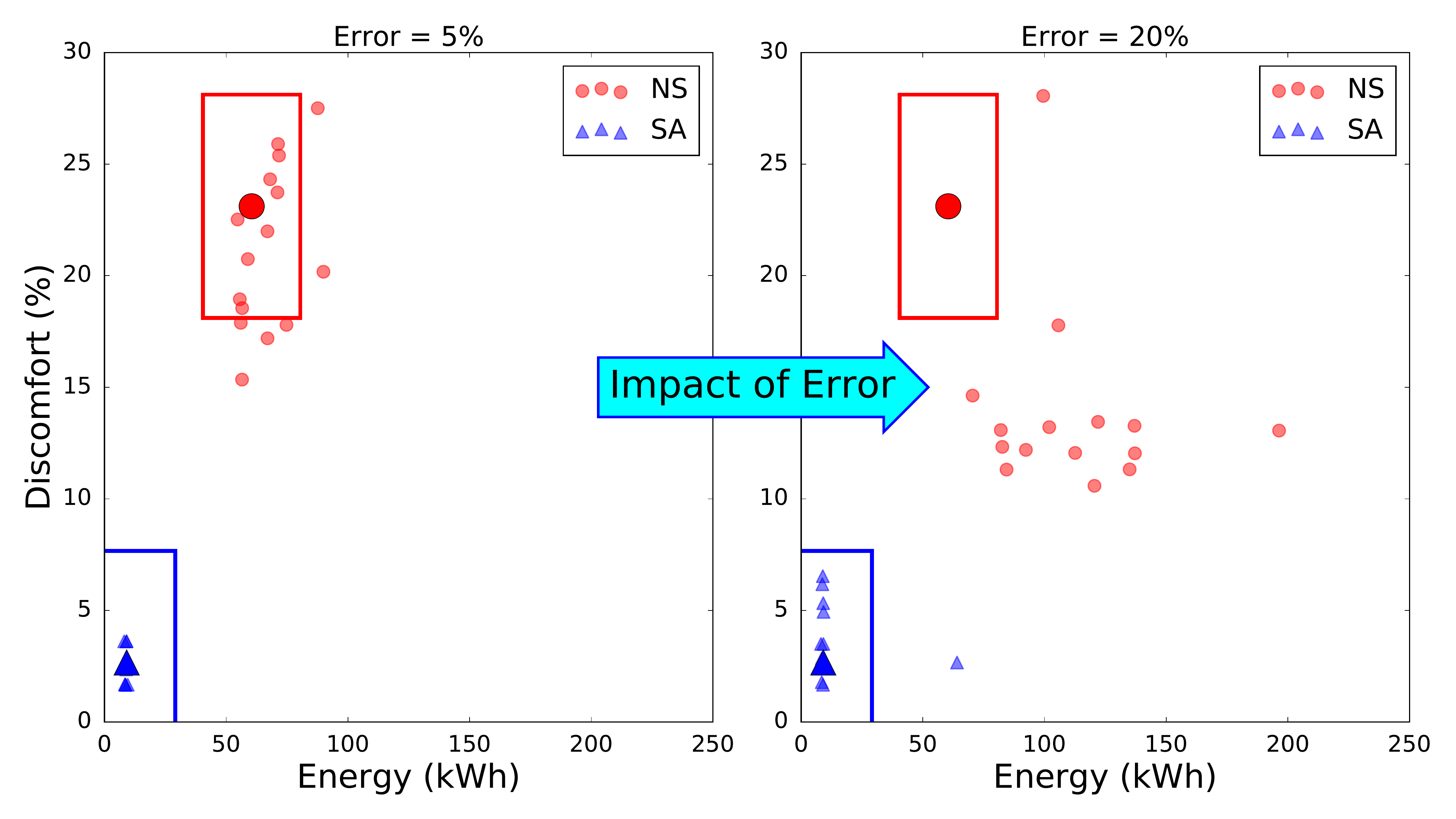}
    \caption{\csentence{Illustration.}
      As error increases, the energy consumption and occupants' discomfort vary depending on the \emph{nature} and the \emph{timing} of prediction errors. 5\% errors on the left and 20\% on the right. Large circles/triangles indicate a perfect prediction scenario and small circles/triangles correspond to those scenarios when occupancy prediction was erroneous.}
    \label{fig:illustration}
\end{figure}

\subsubsection{Robustness}
Prediction errors are stochastic in nature and their impact on energy consumption and occupant comfort depends on two factors:

\textbf{Nature of the Error:} If the prediction algorithm mispredicts occupancy for short time intervals (say for a minute or so), we term the prediction errors as point errors, otherwise we call them burst errors. For a particular error percentage, an erroneous occupancy string can have point errors, burst errors, or a mix of both; resulting in different values of energy consumption and occupants' discomfort for the  \textit{same} error percentage. 

\textbf{Timing of the Error:} The occupancy prediction algorithm can make errors at any time of the day - such as during peak or non-peak time. Consider the situation where the occupancy prediction has 15\% error during the peak hours and the controller assumes one of the five rooms to be occupied though it was unoccupied.
In this situation there is a high chance that the HVAC might be already  running during that time. Given the fact that the other four rooms are occupied, this particular  prediction error will have an insignificant impact on the HVAC operations. However, during night time, the same error percentage might waste significant energy.
This illustrates that the \textit{timing} of the prediction errors has a significant impact on both
comfort and energy consumption. 

For a specific error percentage, depending on the nature and timing of the errors, the energy consumption and user discomfort may either increase or decrease, potentially destabilizing HVAC operations. For a  specific example, consider the big circle and triangle in Figure~\ref{fig:illustration}, which depict the energy consumption and user discomfort for NS and SA controllers respectively for perfect occupancy predictions in a particular simulation scenario. 
For a specific error percentage, the small circles (NS) and triangles (SA) depict the energy consumption and user discomfort for fifteen different erroneous occupancy strings. We noticed that as prediction error increases from $5\%$ (left) to $20\%$ (right), the points indicating erroneous strings start moving away from the results obtained from perfect prediction. 

Note that the circles (NS) are more scattered than the triangles (SA). In the case of NS, the system decides the control parameters such that the desired room temperature (which is the same for each room) is achieved across all the rooms. In case of a sudden change in the occupancy, NS updates the control parameters, but it takes significant time to re-attain the energy-discomfort tradeoff setpoint. In contrast, in SA, the controller knows the current state of SPOT; thus, the controller chooses a set point such that HVAC provides a certain level of comfort to the occupants and SPOT provides the necessary additional offset. SPOT, being responsive in nature, keeps the comfort level of individuals within the desired range with insignificant increase in aggregate energy consumption. Therefore, even if the error percentage increases, the energy and discomfort stays close to the perfect prediction for SA whereas NS becomes highly unstable. 

To capture this phenomenon, Equation~\ref{eq:robust} defines a $robust$ ($cs \in \{NS, SA\}$) metric which quantifies the robustness of a particular control strategy $cs$ towards the prediction errors. It computes the number of instances that stay within the desired limits of the building manager. 

\begin{equation}
	robust_{cs}~(\%) = \frac{\text{\# of instances within limits}}{\text{total \# of instances}} \times 100
	\label{eq:robust}
\end{equation} 

For concreteness,
we use $\pm 20~kWh$  and $\pm 5\%$ as the acceptable limits for energy consumption and occupants' discomfort, respectively, as shown by the rectangles in the figure. For the given scenario (in Figure~\ref{fig:illustration}), when the error percentage is increasing from $5\%$ to $20\%$, NS is less $robust$ towards the prediction error ($60\% \rightarrow 0\%$), however, SA remains consistent ($100\% \rightarrow 93\%$). For a predictive control strategy, a PEC system (like SPOT) mitigates the effect of prediction errors to make the HVAC operations more reliable and robust. Whenever there is an unexpected occupancy in the room, SPOT can react quickly as compared to central HVAC system which has a slower time-scale. 

\section{Evaluation}
	\label{sec:evaluation}
\subsection{Test Building Description}
\begin{figure}[h!]
	\includegraphics[width=0.8\linewidth]{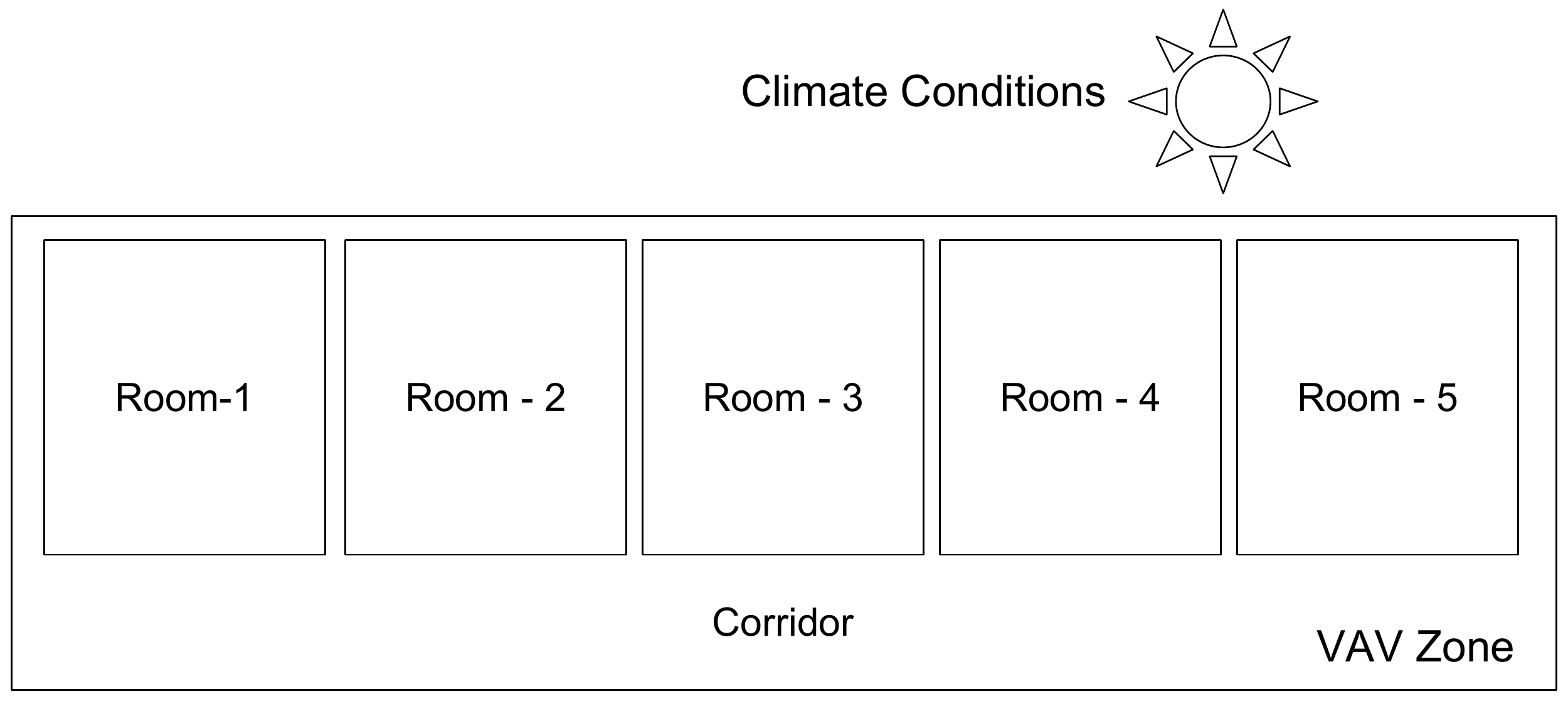}
    \caption{\csentence{Setup.}
      For evaluation, we considered a hypothetical building consisting of 5 rooms separated by walls.}
    \label{fig:setup}
\end{figure}

For our evaluation, we consider a single zone in a typical building comprising of five rooms where each room is surrounded by walls on three sides and has a window exposed to weather conditions on the fourth (Figure~\ref{fig:setup}). We assume an AHU and a VAV unit in the building. Though the structure is hypothetical, it is a typical architecture for faculty offices in  Universities where thick brick walls separate the rooms. We believe that the key insights obtained for the study are well representative of more complicated building architectures. Note that ThermalSim can also deal with more complicated structures, should that be desired. 
When we evaluate MPC with a reactive controller, we also consider effect of SPOT heater/fan on the room temperature. For the stated scenario, we next discuss the dataset.

\subsection{Dataset}
ThermalSim requires  real-world occupancy data to generate an \emph{error matrix}. We leveraged an existing deployment from our university campus and gathered occupancy data (along with other information) from more than fifty volunteers -- including students, faculty, and the staff members every 30 seconds for a year.

\begin{figure}[h!]
	\includegraphics[width=0.9\linewidth]{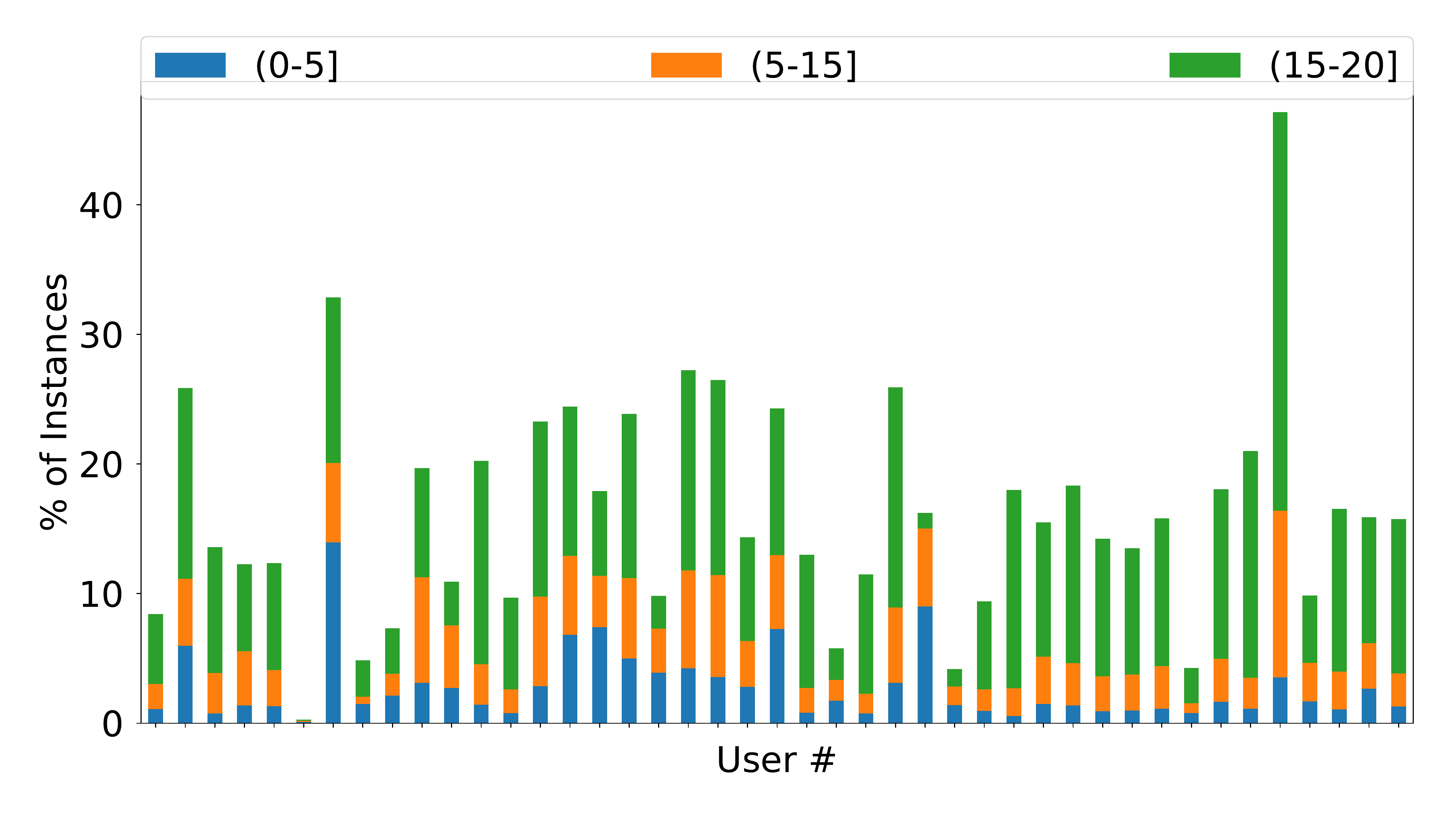}
    \caption{\csentence{Variation in Occupants' Schedule.}
      We sample occupancy every 30 seconds; in every 10-minute interval, there exist 20 measurements of occupancy information. Here, the color indicates the number of 30 seconds instances in a 10-minute interval when the room was occupied. Notice that room would be marked occupied for all the three scenarios, however, the percentage of instances when the room was occupied for less than 2 minutes (in the range of (0, 5]) is relatively low.}
    \label{fig:variation}
\end{figure}

\begin{figure}[h!]
	\begin{minipage}[t]{0.44\linewidth}
		\includegraphics[width=0.9\linewidth]{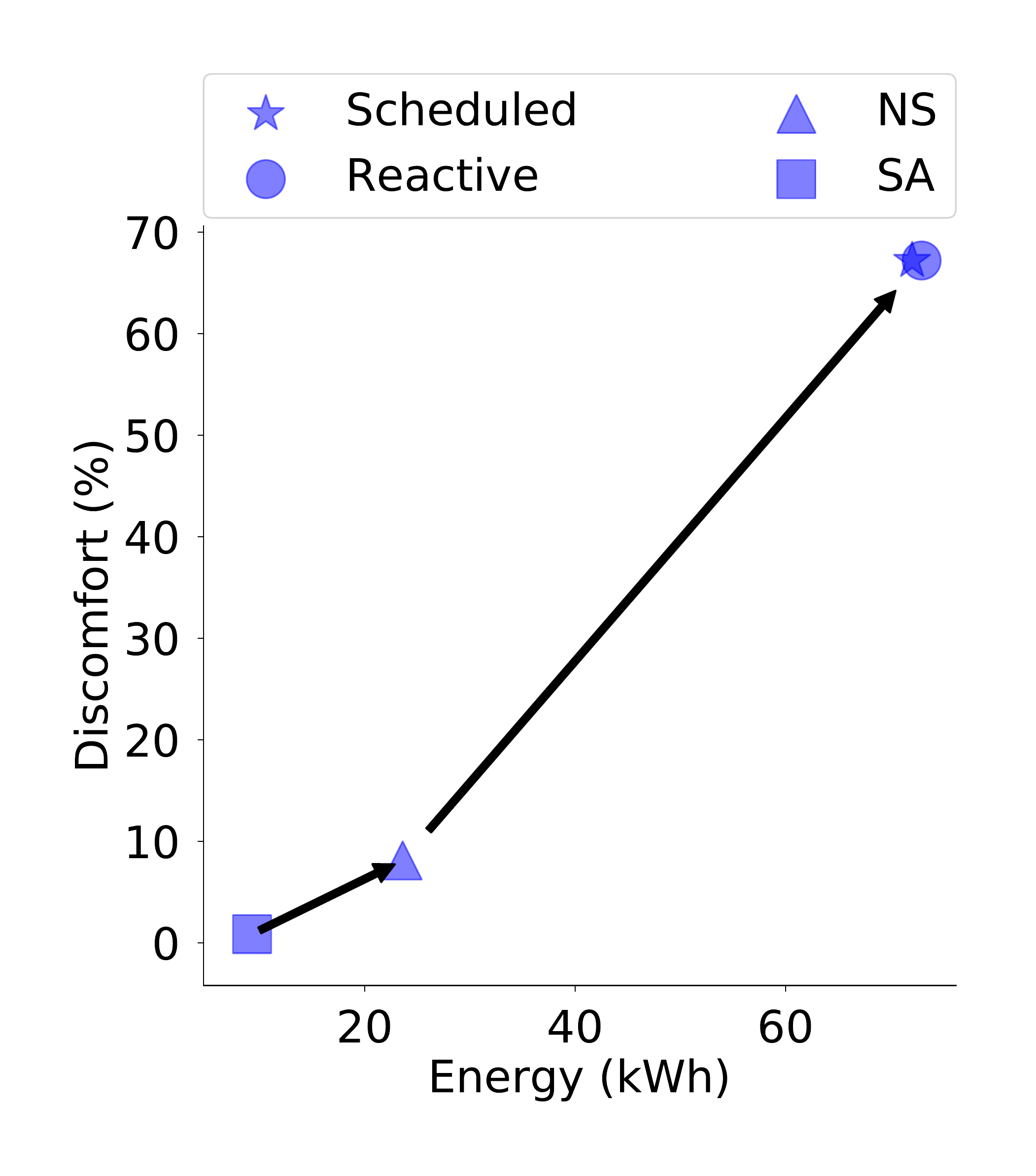}
    		\caption{\csentence{Summers.} Energy-discomfort plot when prediction is perfect. The arrow indicates the performance degradation, in terms of energy consumption and user comfort, when we move from predictive to non-predictive control strategies.}
   	\label{fig:day_com_scatt}
    \end{minipage}
	\begin{minipage}[t]{0.5\linewidth}
		\includegraphics[width=0.9\linewidth]{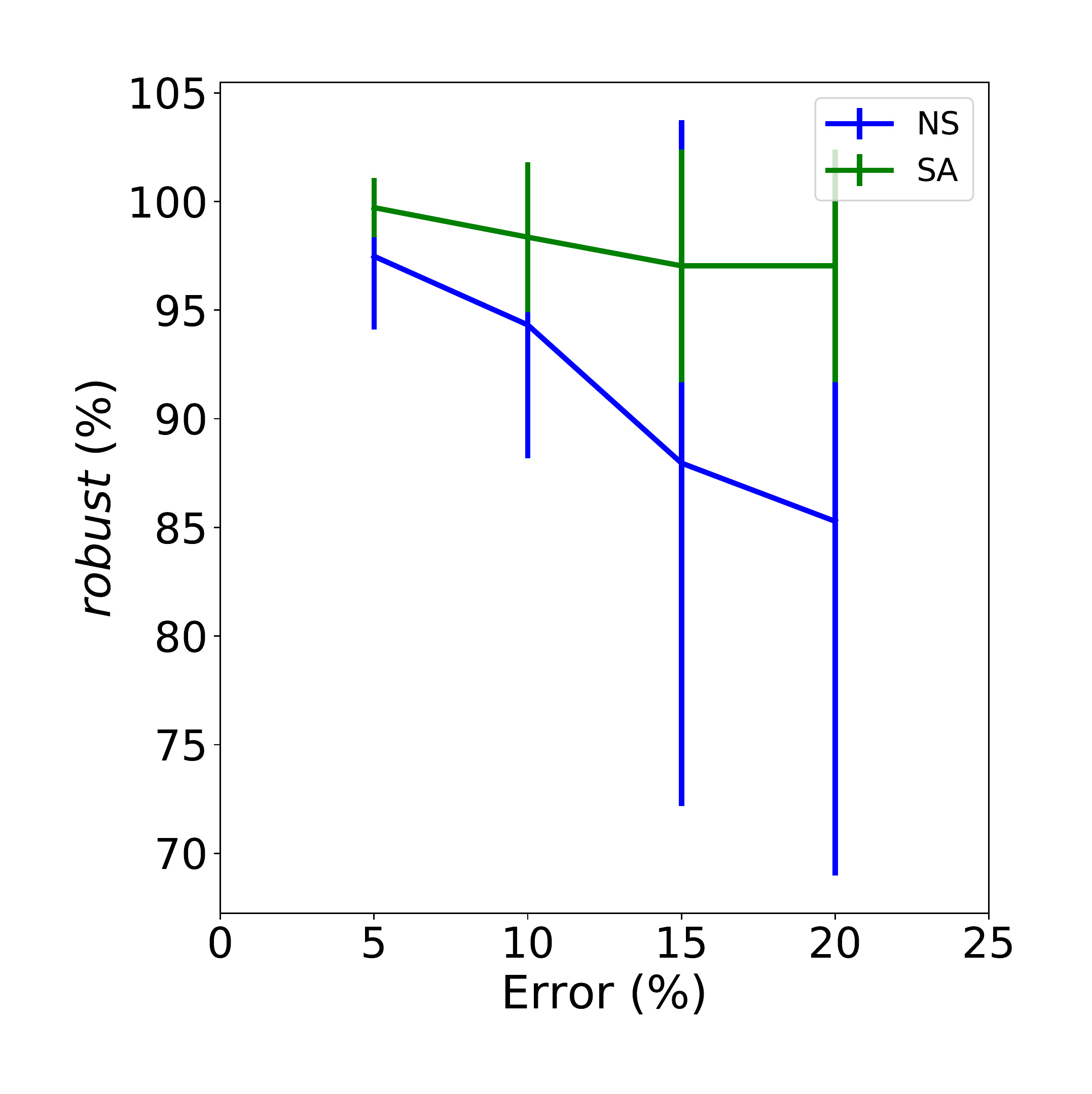}
    		\caption{\csentence{Summers.}
      As error increases $5\% \rightarrow 20\%$, SA stays more robust than NS. Error bars indicate the variation in different simulated scenarios. For system to be more robust, the length of error bar should be smaller.}
    		\label{fig:err_wise_csummers}
    \end{minipage}
\end{figure}

\begin{figure}[h!]
	\includegraphics[width=0.9\linewidth]{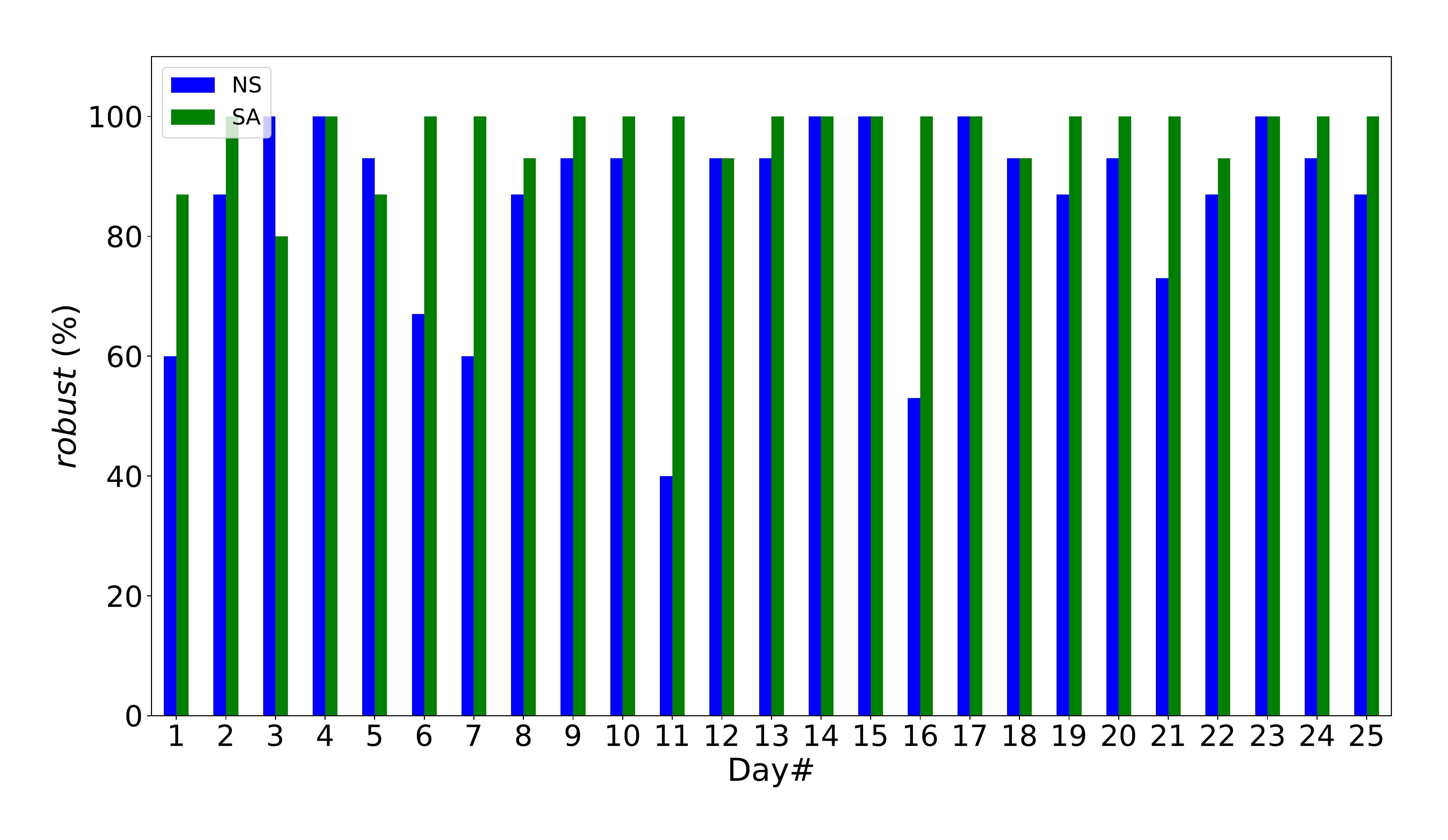}
	\caption{\csentence{Summers.} For $20\%$ prediction error in occupancy, SA is more reliable and robust NS across all the 25 days of summer.}
	\label{fig:day_wise_csummers}
\end{figure}

An MPC requires occupancy information in every 10-minute to compute the control parameters 24 hour time horizon; therefore, we upsample the occupancy data from 30-seconds to 10-minutes by applying the following rule - \textit{``Mark a 10-minute interval unoccupied if all the 30-second instances indicate the room to be unoccupied, else mark the space as occupied''.} However, with this strategy, even if a single instance in the 10-minute interval is occupied, the controller will mark the space as occupied for the whole duration. To understand whether such a bias is limiting or not, we analyzed the occupancy data and our analysis indicates that data has only 3\%  10-minute instances where the room is occupied for less than 2 minutes (Figure~\ref{fig:variation}). Therefore, we only mark a 10-minute interval unoccupied if the room was occupied at all the 30-second instances within that interval.

\subsection{Evaluation Setup}

Our research hypothesis is that 
the benefits of using a PEC system like SPOT along with HVAC controller
mitigates the influence of prediction errors on MPC-based HVAC operation.

We validate this hypothesis assuming occupants in all the five rooms have similar comfort requirements: $[23^\circ$C$, 25^\circ$C$]$ in summers and $[21^\circ$C$, 23^\circ$C$]$ in winters. 
For the given setup, we compare the performance of predictive 
and non-predictive HVAC controllers for 25 days, both in summers and winters.

\begin{figure}[h!]
\begin{minipage}[t]{0.44\linewidth}
\includegraphics[width=0.9\linewidth]{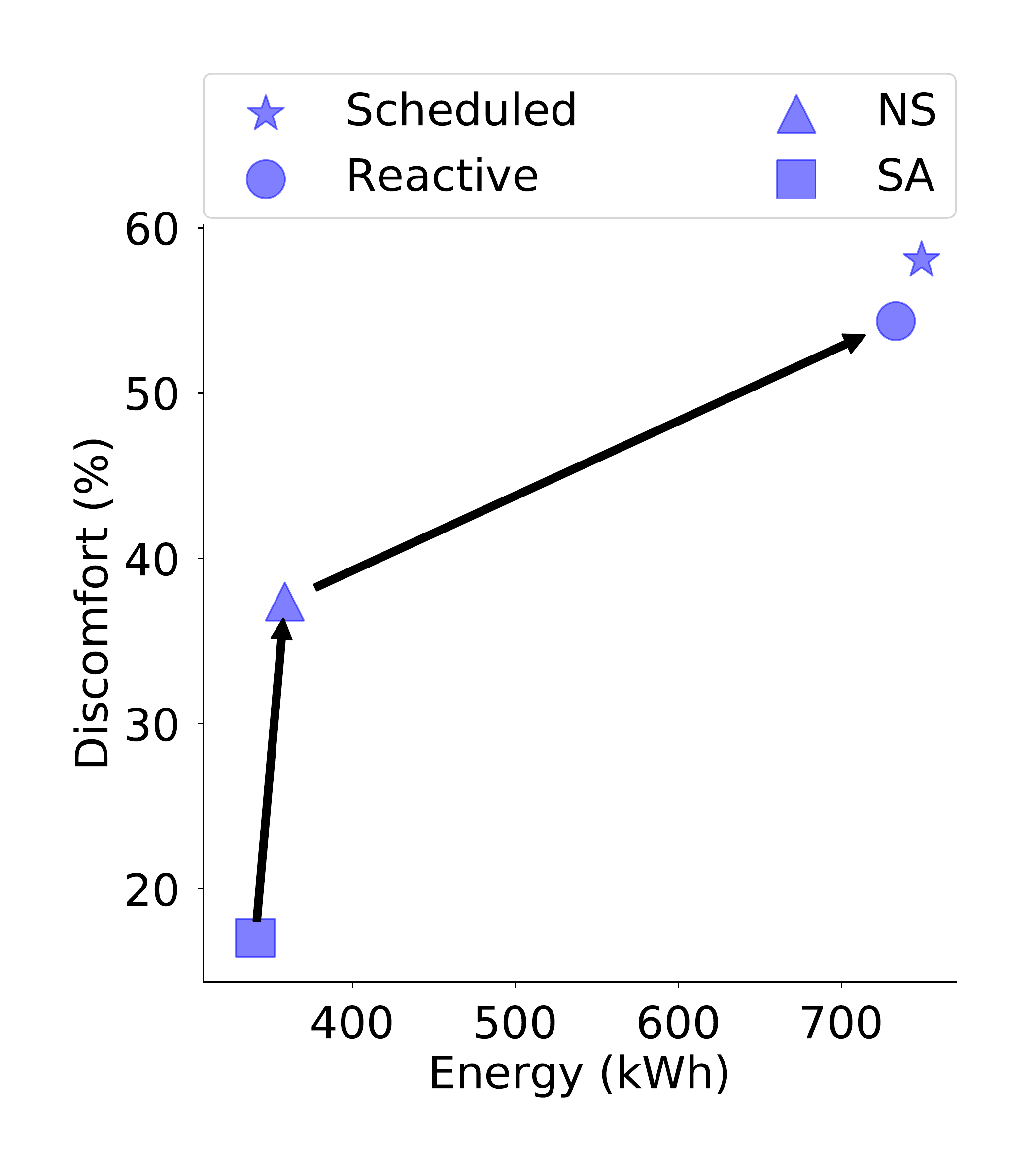}
    \caption{\csentence{Winters.}
      Energy-discomfort plot when prediction is perfect. The arrow indicates the performance degradation, in terms of energy consumption and user comfort, when we move from predictive to non-predictive control strategies.}
    \label{fig:wday_com_scatt}
\end{minipage}
\begin{minipage}[t]{0.5\linewidth}
\includegraphics[width=0.9\linewidth]{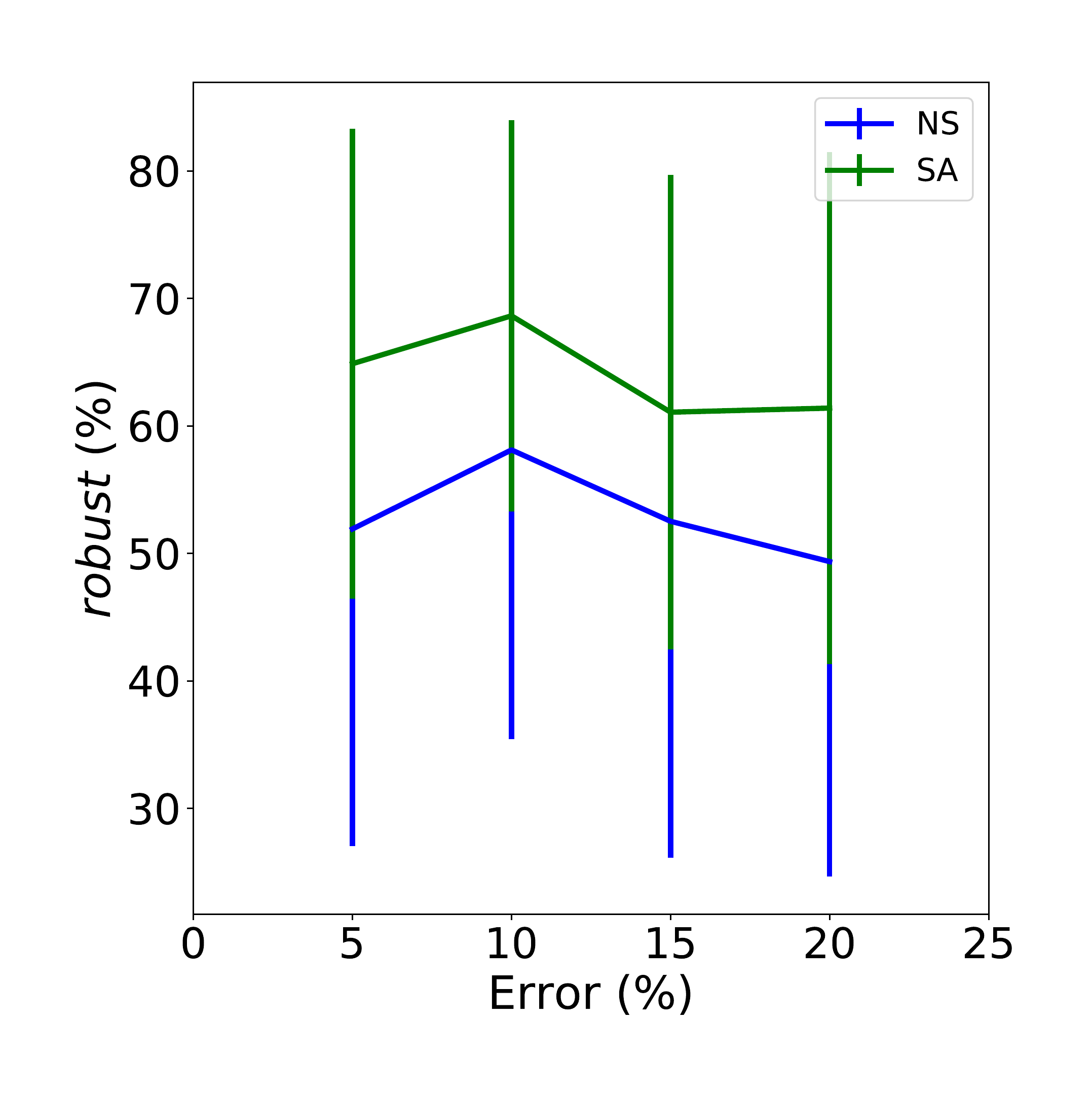}
    \caption{\csentence{Winters.}
      Even with slow heater, SA is better or comparable than NS. Error bars indicate the variation in different simulated scenarios. For system to be more robust, the length of error bar should be smaller.}
    \label{fig:err_wise_cwinters}
\end{minipage}
\end{figure}

\begin{figure}[h!]
\includegraphics[width=0.9\linewidth]{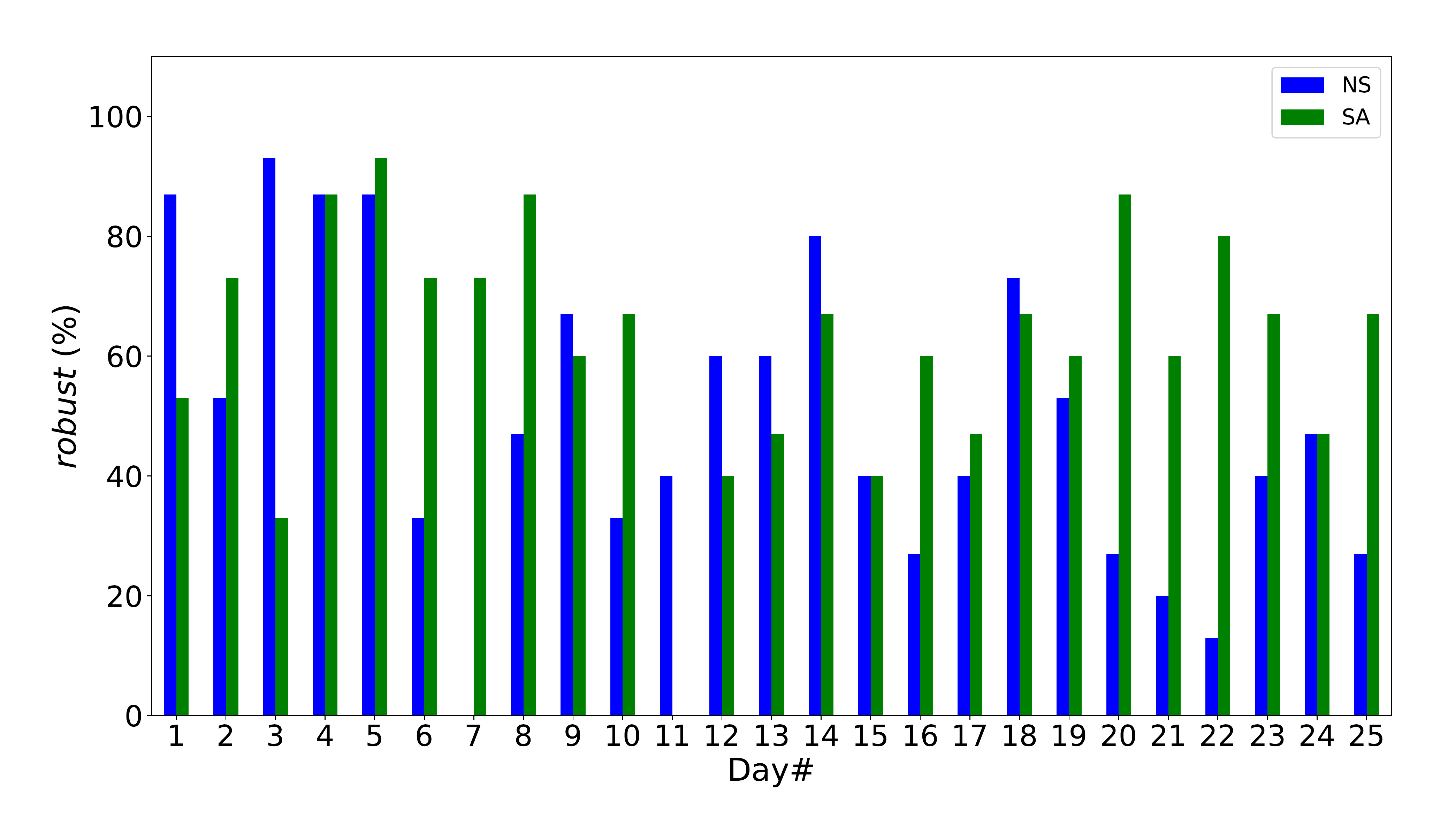}
    \caption{\csentence{Winters.}
      For error percentage as high as $20\%$, note that SA has less deviation in HVAC operations than NS.}
    \label{fig:day_wise_cwinters}
\end{figure}

For each day, we select an occupancy string from the \emph{error matrix} that deviates (from the current day) by the error percentage specified in the system. For instance, if we wish to introduce 10\% error in the current day occupancy string, we search for another occupancy string in historical data where 288 out of 2880 instances (for a data sampled every 30 seconds) have a mismatch with the current day occupancy string. \emph{ThermalSim} utilizes both actual and erroneous occupancy string to simulate the building (depicted in Figure~\ref{fig:setup}) for all the four control strategies and compare their performance. 

To mitigate any bias in the selection of erroneous occupancy strings, \emph{ThermalSim} evaluates fifteen different erroneous occupancy patterns for each day and error percentage. Furthermore, a separate analysis for each of the two  seasons provides better understanding of the influence of seasonal variations. 

\subsection{Insights \textcolor{white}{updated the title and text of this whole section}} 

\label{sec:sim}
In the jurisdiction corresponding to 
our temperature data set, i.e., Southwestern Ontario, 
we find that for all control strategies,
the HVAC system consumes less energy in summers when compared to winters (see Figures~\ref{fig:day_com_scatt}, \ref{fig:wday_com_scatt}). 
In our setting, the outside temperature in summers is only a few degrees higher than the desired room temperature and so the HVAC has to put in
less effort to achieve the desired comfort. On the other hand, in winters, the HVAC energy consumption is significantly higher because the outside 
temperature is quite cold. In winters, all control strategies attempt
to maintain a room temperature in the range of $21^\circ C$ and $23^\circ C$ which is much higher than the outside temperature (approx. $-10^\circ C$). Consequently, HVAC has to expend more energy in winters than in summers
to attain the desirable comfort conditions in the occupied zones. 

\subsubsection*{User Experience}
The schedule-based and reactive controllers can make occupants uncomfortable and yet consume significant energy, even with perfect prediction. When set to follow a fixed schedule, HVAC supplies air at a constant flow and temperature, and does not consider occupants' schedules or
daily temperature changes. For pictorial representation, we use energy-discomfort plot where x-axis denotes the daily energy consumption of the building and y-axis represents the total discomfort for the users.
Consequently, with a \textit{schedule-based} control strategy,
user experience lies in the top-right corner of the energy-discomfort plot with maximum energy consumption along with notable discomfort for both the seasons (see Figures~\ref{fig:day_com_scatt}, \ref{fig:wday_com_scatt}). On the other hand, a reactive controller with occupancy information is marginally better or equivalent to the schedule-based controller. Model predictive control (with no SPOT) shows significant improvement in minimizing both energy consumption and occupants' discomfort. Given the weather forecast and occupancy prediction, MPC keeps updating the temperature and volume of supply air at regular time intervals. 

As central HVAC unit cannot cater to the dynamic schedule of the occupants, discomfort in NS is slightly higher than the hybrid control strategy that integrates SPOT with MPC to satisfy the comfort requirements of each individual in the building. In SA, the central HVAC system is aware of the SPOT system, therefore, the controller choses the set point temperature such that HVAC can provide minimal comfort, and SPOT can offset the individual comfort requirements. This results in additional savings in energy when there is partial occupancy is in line with the results from previous study by Rachel et al.~\cite{Rac17}. Next, we observed that the discomfort is negligible for summer as opposed to winter. The fan assists the occupant in quickly achieving her desired comfort level as opposed to a heater which takes comparatively more time to increase the temperature to provide the offset. In conclusion, irrespective of the season, both SA and NS strategies improve comfort and energy compared to schedule-based and reactive, with SA outperforming NS.  

\subsubsection*{Error Analysis}
When occupancy predictions are erroneous, depending upon the \emph{nature} and \emph{timing} of errors, energy consumption and occupants' discomfort vary, hence HVAC operations become highly variable. When analysed over 25 days each for 15 different occupancy patterns, we find that the SA control strategy is more robust than NS even with a high error percentage. As the prediction error increase from $5\%$ to $20\%$, the performance of NS  drops while SA performance remains quite consistent (Figure~\ref{fig:err_wise_csummers}). For $20\%$ prediction error, SA ($\sigma=5\%$) is $12\%$ more robust than NS ($\sigma=16\%$).

Next, Figure~\ref{fig:day_wise_csummers} shows that the SA is consistently robust across all the days as compared to NS for $20\%$ prediction error in the occupancy. Highly unreliable HVAC operations lead to significant variations in the energy consumption and the user comfort. Though the fan makes insignificant impact on the room temperature, it quickly achieves the desired comfort level by providing a cooling perception to the user. Thus, the fan is very helpful in dealing with the unexpected changes in the occupancy of the room while NS alone fails to do so. 

We noticed that a fan is more effective and quicker than a heater in mitigating the effect of prediction errors on both energy consumption and discomfort. When a room gets occupied, a heater slowly increases the room temperature to achieve the comfort requirements of the occupants. This results in few intervals of discomfort for the user. This effect is visible in Figures~\ref{fig:err_wise_cwinters}, \ref{fig:day_wise_cwinters}. For $20\%$ prediction error, we noticed that the SA is now even less robust than NS on few days due to the slow reaction of the SPOT heater. However, the average performance of SA is still better or comparable than NS.

\section{Discussion and Conclusion}
	\label{sec:discussions}
In this work, we analysed the influence of prediction errors in occupancy on the HVAC operations while leveraging a custom-built building simulator - \emph{ThermalSim}. In this section, we summarize our results, discuss various limitations of the study followed by research questions which are open for the community. 

Our insights include the following: First, our dataset indicates that aggregate energy consumption is higher in winters than in summers. Second, integrating a PEC like SPOT with a predictive HVAC controller is definitely better or comparable than a pure MPC based approach. Third, for SA controller, fast reactive device (such as fan) is $20\%$ better than the heater, in terms of occupants discomfort. Finally, NS typically fails to satisfy the comfort requirements on any day. 

Our work suffers from two main limitations.
First, while the thermal model (of \emph{ThermalSim}) considered the effect of numerous sources (such as weather, occupancy) affecting the room temperature, there still exist various other factors (such as humidity) which are critical for such analysis. We plan to explore  such factors and enrich the data for a deeper analysis in future. 
 
Second, we carried out the study through a dataset collected from a particular part of the world. Climate, users' attitude (towards energy savings), and many other factors differ significantly across the geographies. Though the results indicate that SA is more robust than NS, there can be considerable discrepancy across (and within) the countries. A real-world implementation of the technology is critical to understand its effectiveness in achieving the desired goals.

We find that mitigating the effect of prediction errors possess considerable potential in optimising the HVAC operations with predictive controllers. While model predictive control (MPC) is one of the most promising state of the art HVAC control strategies, its performance is limited by the accuracy of the weather and occupancy predictions. Therefore, we designed a custom-built building simulator -- \emph{ThermalSim} -- to analyse the influence of prediction errors on HVAC operations. We also proposed a  method to introduce realistic errors in occupancy for the analysis. Our initial analysis indicates that prediction error (in occupancy) of $20\%$ can make the HVAC operations highly unstable in terms of both energy consumption and occupants' comfort. Recent literature shows that it is feasible to use a personal thermal comfort system -- SPOT -- along with predictive strategy to ensure personalised comfort in personal and shared spaces. We observed that while SPOT is effective in attaining better personalised comfort, it also strengthens the predictive strategies by mitigating the influence of predictions errors on energy consumption and occupants' comfort because it works at a finer time-scale than the MPC-based HVAC. Employing a personal thermal comfort system, such as SPOT, we stay in the acceptable region 95\% of the times as oppose to 83\% of the times even for the prediction errors as high as $20\%$, in the occupancy; thus, motivating a reliable control strategy across the commercial buildings. 


\begin{backmatter}
\section*{Availability of data and materials}
    On request, the authors will supply occupancy data to interested researchers. The simulator is open source and available at \url{https://github.com/milanjain81/SBS_MakefileProject}.

\section*{Competing interests}
    The author(s) declare(s) that they have no competing interests.

\section*{Funding}
    This work was supported by Cisco Systems Canada and the Canadian Natural Sciences and Engineering Research Council under a Cooperative Research and Development grant. The funding body played no role in the design of the study, collection, analysis, and interpretation of data, and in writing the manuscript.

\section*{Author's contributions}
    Together, all the authors devised the project, the main conceptual ideas and outline for the analysis. Milan Jain carried out the implementation and analysed the data. Prof. Kalaimani,  Prof. Keshav, and Prof. Rosenberg worked out the optimisation problem for Model Predictive Control and helped in writing the manuscript. 

\section*{Acknowledgements}
    We would like to acknowledge Alimohammad Rabbani (then a Masters student at University of Waterloo) and Costin Ograda-Bratu (currently a lab technician at University of Waterloo) for collecting and sharing occpancy data. We would also like to acknowledge the volunteers who participated (and are even now participating) in data collection. 


\bibliographystyle{bmc-mathphys} 
\bibliography{error18_ei}      







\end{backmatter}
\end{document}